\documentclass[%
 reprint,
 superscriptaddress,
 amsmath,amssymb,
 aps,
]{revtex4-2}

\usepackage{graphicx}
\usepackage{dcolumn}
\usepackage{bm}
\usepackage{hyperref}
\usepackage[colorinlistoftodos]{todonotes}

\newcommand{\msr}{$\mu$SR}
\newcommand{\avs}{$A$V$_3$Sb$_5$}
\newcommand{\cvs}{CsV$_3$Sb$_5$}
\newcommand{\kvs}{KV$_3$Sb$_5$}
\newcommand{\cvts}{Cs(V$_{0.95}$Ti$_{0.05}$)$_3$Sb$_5$}

\newcommand{\svo}{Sr$_{2}$VO$_{4}$}

\newcommand{\tcr}[1]{\textcolor{black}{#1}} 

\begin{document}

\title{Prevailing orbital excitations in paramagnetic kagome superconductor \cvts}
\author{Chennan Wang}
\email{chennan.wang@unifr.ch}
\thanks{These authors contributed equally}
\affiliation{Department of Physics, University of Fribourg, Fribourg, CH-1700, Switzerland}
\affiliation{PSI Center for Neutron and Muon Sciences, Paul Scherrer Institut, Villigen, CH-5234, Switzerland}

\author{Yuhang Zhang}
\thanks{These authors contributed equally}
\affiliation{Beijing National Center for Condensed Matter Physics and Institute of Physics, Chinese Academy of Sciences, Beijing 100190, China}

\author{Zhen Zhao}
\affiliation{Beijing National Center for Condensed Matter Physics and Institute of Physics, Chinese Academy of Sciences, Beijing 100190, China}

\author{Zhouyouwei Lu}
\affiliation{Beijing National Center for Condensed Matter Physics and Institute of Physics, Chinese Academy of Sciences, Beijing 100190, China}

\author{Hui Chen}
\affiliation{Beijing National Center for Condensed Matter Physics and Institute of Physics, Chinese Academy of Sciences, Beijing 100190, China}

\author{Ziqiang Wang}
\affiliation{Department of Physics, Boston College, Chestnut Hill, Massachusetts 02467, USA}

\author{Haitao Yang}
\affiliation{Beijing National Center for Condensed Matter Physics and Institute of Physics, Chinese Academy of Sciences, Beijing 100190, China}

\author{Christian Bernhard}
\affiliation{Department of Physics, University of Fribourg, Fribourg, CH-1700, Switzerland}

\author{Xiaoli Dong}
\affiliation{Beijing National Center for Condensed Matter Physics and Institute of Physics, Chinese Academy of Sciences, Beijing 100190, China}

\author{Hong-Jun Gao}
\affiliation{Beijing National Center for Condensed Matter Physics and Institute of Physics, Chinese Academy of Sciences, Beijing 100190, China}


\begin{abstract}

Using the muon as a sensitive local magnetic probe, we investigated the layered kagome superconductor \cvts, a material notably devoid of both static magnetic moments and long-range charge order. Our transverse-field \msr~measurements reveal that the local magnetic susceptibility, obtained via the muon Knight shift, is dominated by orbital excitations with a split energy levels around 20 meV. Meanwhile, the persistence of itinerant electron paramagnetism down to 5 K and 7 T confirms the absence of static magnetism within this regime. In addition, zero-field (ZF) \msr~experiments detect a significant increase in the inhomogeneous nuclear dipolar field distribution below a featured temperature at 70 K. We attribute this ZF-\msr~feature to the emergence of local lattice distortions at low temperatures, potentially arising from orbital ordering. Significantly, our study establishes that orbital excitations constitute an intrinsic property of the layered V-Sb kagome lattice. Despite its small magnitude, spin-orbit coupling plays a crucial role in governing the lattice dynamics, potentially driving the emergence of novel phenomena such as phonon carrying angular momentum in crystals with non-chiral point groups.
\end{abstract}

\maketitle
\textit{Introduction- }Kagome compounds are materials that feature at least one atomic sublattice arranged in a kagome geometry, a pattern of corner-shared triangles. This structure inherently induces strong geometric frustrations in the coupled spin and charge degrees of freedom \cite{ong_electronic_2004}. The vanadium based kagome superconductors \avs~(A = K, Rb, Cs) represent the first family of two-dimensional kagome metals \cite{wilson_av3sb5_2024} and have attracted intense interest due a cascade of intertwined electronic phases with exotic properties. For the \cvs~compound, numerous phenomena emerging below its 94 K transition into a long-range 2×2 charge density wave (CDW) state remain enigmatic. These include a unidirectional charge order that breaks rotational symmetry \cite{li_unidirectional_2023} —distinct from nematic instabilities preserving translational symmetry \cite{liu_absence_2024}, anomalous $c$-axis nonreciprocal transport \cite{guo_switchable_2022}, superconducting diode effects \cite{le_superconducting_2024,ge_nonreciprocal_2025}, superconductivity exhibiting broken rotational symmetry \cite{ni_anisotropic_2021}, and, remarkably, a CDW state that spontaneously breaks spatial inversion symmetry, manifesting intrinsic chirality \cite{jiang_unconventional_2021,shumiya_intrinsic_2021,wang_electronic_2021}. Elucidating the origins and interplay of these phenomena presents an ongoing challenge.

Despite of a non-chiral structure of the \avs~(P6/mmm), the chiral CDW in \avs~compounds was first proposed based on scanning tunneling microscopy (STM) observations \cite{jiang_unconventional_2021}. The justification stems from the Fourier-transform patterns exhibiting circular intensity modulations with a definite handedness, which is reversibly switched by the polarity of an applied out-of-plane magnetic field \cite{xing_optical_2024}. Further support for chirality in the electronic structure of the parent compound \cvs~has also been reported \cite{elmers_chirality_2025}.

Theoretical proposals suggest that CDW formation on the kagome lattice can generate charge loops, potentially breaking time-reversal symmetry \cite{zhou_chern_2022}. Nevertheless, the microscopic origin of the chiral CDW remains unresolved. Two primary scenarios are under active consideration: (1) An electronically driven mechanism, where orbital magnetism emerges from loop-current order, although the currents themselves need not be chiral \cite{zhou_chern_2022}; and (2) A phonon-driven mechanism, mediated by chiral phonons carrying finite angular momentum from ionic displacements (e.g. circularly polarized lattice vibrations) \cite{zhang_understanding_2024}, which is predicted to exist in kagome lattices \cite{zhang_angular_2014,chen_chiral_2019}. For the latter scenario, experimental observations point to electron-phonon coupling playing a dominant role over purely electronic interactions in CDW formation. This is manifested by significant band renormalization and kink features near the Fermi level \cite{luo_electronic_2022}, suggesting a strong coupling to the lattice.

While phonons are fundamentally linearly polarized lattice vibrations, they can acquire angular momentum by coupling to orbital excitations within the system. These orbital excitations may originate from strong electronic instabilities that induce lattice distortions or, alternatively, from spin-orbit coupling, an intrinsic atomic-scale mechanism independent of strong electronic correlations. However, direct experimental identification of such orbital excitations remains elusive in the \avs~compounds, and consequently, understanding their precise role is critical. Experimentally, chiral phonons can manifest magnetic-field-dependent responses \cite{baydin_magnetic_2022,ohe_chirality-induced_2024}. These responses arise from the interaction between the applied field and the effective magnetic moments intrinsic to chiral phonon modes \cite{juraschek_orbital_2019}. This interaction, known as the phonon Zeeman effect, is mediated by orbit-phonon coupling \cite{mclellan_angular_1988,juraschek_giant_2022,chaudhary_giant_2024} and perturbs local atomic displacements under an applied magnetic field.

Here, we utilize muon spin rotation (\msr) to gain deeper insights into the local magnetism and structural distortions (via the presented local nuclear fields) in this compound. Muons, once implanted, occupy interatomic lattice sites, acting as sensitive probes of their immediate environment through interactions with local spin susceptibility and magnetic fields \cite{hillier_muon_2022}. In an applied magnetic field, the field-induced signatures detected via \msr~predominantly arise from induced magnetic moments, alteration of the local spin susceptibility, or local distortions within the crystal lattice proximal to the muon site, allowing for detailed inspection at a localized scale.

\textit{Paramagnetic susceptibility- }The Ti doped ($x=0.05$) \cvts~single crystals were prepared as described in Ref. \cite{yang_titanium_2022}. The doped compound maintains the same point group symmetry as the original structure [Fig. 1(a)] and possesses a nearly identical lattice parameter, with only a slightly shorter $c$-axis lattice parameter \cite{yang_titanium_2022}. Magnetization and transport measurements confirm the absence of long-range charge order [Fig. 1(b)] while preserving bulk superconductivity with a transition temperature $T_c=3$ K \cite{yang_titanium_2022}. This establishes Ti doping as an effective pathway to suppress the CDW state without destroying superconductivity.

Crucially, the Ti-doped compound maintains paramagnetic behavior without any long-range magnetic order down to 5 K [Fig. 1(c)]. Magnetization measurements show Curie-Weiss behavior with a weak temperature dependence between $100–300$ K. At lower temperatures, a distinct deviation arises, which can be attributed to excitations between spin-orbit coupled $\Gamma_7$ (doublet) and $\Gamma_8$ (quartet) states in the ideal octahedral crystal field. Under an applied magnetic field ($H{\parallel}c$), the magnetic susceptibility shows no induced magnetism up to 7 T. The magnetic susceptibility curves can be rescaled by considering the magnetization contribution from Pauli paramagnetism of the itinerant carriers and van Vleck paramagnetism arising from the low-lying multiplet manifold of the V $d$-orbitals.

\begin{figure}[h]
    \centering
    \includegraphics[width=0.95\linewidth]{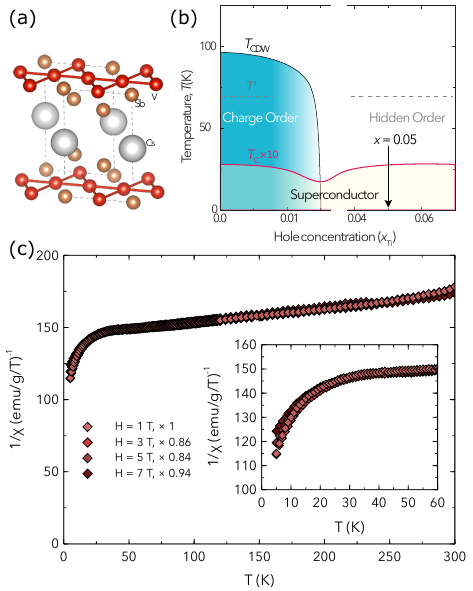} 
    \caption{\label{fig:fig1} (a) Schematic representation of the atomic structure of the \cvts~crystal, with atoms labeled; (b) Electronic phase diagram of \cvs~as a function of Ti doping adapted from Ref. \cite{yang_titanium_2022, Huang_Revealing_2025}; (c) Inverse magnetic susceptibility of Ti doped ($x=0.05$) \cvts~measured on a single crystal. The compound displays a Curie-Weiss behavior with the addition of paramagnetic contributions at higher temperatures. Insets: Low-temperature susceptibility data. The deviation from Curie-Weiss behavior indicates the impact of the crystal electric field on the magnetic susceptibility. Note, for comparison the magnetic susceptibility data for different applied magnetic fields have been multiplied by different scaling factors.
    }
\end{figure}

\textit{Absence of local magnetic moment of electron spin- }The absence of any static and dynamic magnetism in 5\% doped compound is further confirmed by weak transverse field (wTF) and longitudinal field (LF) \msr~experiments. The wTF-\msr~spectra at 5 K [Fig. 2(a)] exhibt no missing asymmetry, ruling out fast depolarization from electronic moments. Meanwhile, the LF-\msr~spectra [Fig. 2(b)] display characteristic decoupling of nuclear dipole fields with increasing field $B_\mathrm{LF} = \mathrm{\omega/\gamma_\mu}$ ($\gamma_\mu/2\pi = 135.5$\,MHz/T), applied parallel to both the $c$-axis and initial muon spin direction. This behavior quantitatively described by the depolarization function as depicted in Equ. 1, which includes dynamic processes including fluctuating moments and thermally activated or quantum tunneling-induced muon hopping \cite{hayano_zero-and_1979}.
\begin{equation}
\begin{split}
P_\mathrm{LF}=1-\frac{2\mathrm{\Delta}^2}{\mathrm{\omega}^2}\left[1-\mathrm{exp}(-\frac{1}{2}\mathrm{\Delta}^2t^2)\mathrm{cos}(\omega{t})\right]\\
    +\frac{2\mathrm{\Delta}^4}{\mathrm{\omega}^3}\int_{0}^{\tau}\mathrm{exp}(-\frac{1}{2}\mathrm{\Delta}^2\tau^2)\mathrm{sin}(\omega\tau)d\tau.
\end{split}
\end{equation}
Therefore, no static or dynamic magnetic moment is evident in the 5\% Ti-doped sample. The muon spin depolarization is primarily due to coupling with the surrounding nuclear moments. These results also suggest that the muon diffusion effect, which can introduce additional dynamical relaxation, is excluded in this case.

\begin{figure*}[ht]
    \centering
    \includegraphics[width=1\linewidth]{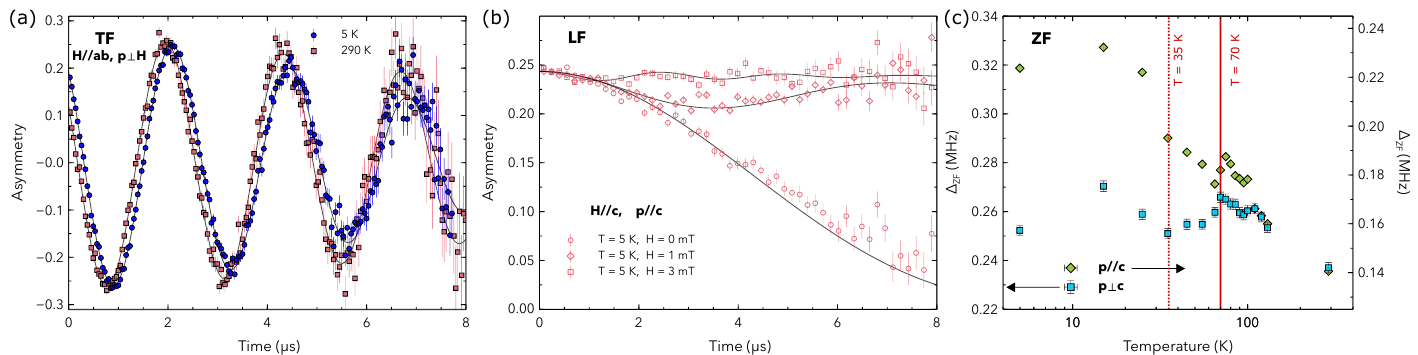}
    \caption{\label{fig:fig2} \msr~spectra taken from Ti ($x=0.05$) doped \cvts~single crystals: (a) in a weak applied transverse field (wTF) of 30 Oe at 5 K and 290 K. Note: a small initial phase shift of the spectra between the two measurements in (a) is due to the different initial muon spin angle settings; (b) in a varies of small applied longitudinal fields (LF) at 5 K. Solid lines are fits to the experimental data as described in the text; (c) ZF-\msr~spectra of the Ti ($x=0.05$) doped sample. The experimental data marked in green and blue are the measured muon spin depolarization rate along the muon spin projection direction within the crystal $ab-$plane and $c-$axis, respectively. 
    }
\end{figure*}

\textit{Anisotropic muon spin depolarization in zero-field (ZF)- }In a previous ZF-\msr~experiment on the \cvs~parent compound, an anomalous anisotropic feature of the local magnetic field distribution was discovered \cite{yu_evidence_2021}. Its temperature dependence indicates that a hidden state emerges in the bulk below 70 K, a temperature distinct from the CDW onset temperature.

For the Ti doped compound, we performed ZF-\msr~experiments using the same measurement configurations as described in Ref. \cite{yu_evidence_2021}. As shown in Fig. 2(b), nuclear moments dominate the muon spin depolarization, which can be phenomenologically described by a Gaussian Kubo-Toyabe (GKT) function \cite{hayano_zero-and_1979} for $H = 0$ mT. It is worth noting that a slight fitting deviation from the GKT function is expected, as its rigorous application requires densely distributed and isotropic nuclear moments, a condition violated in this layered material \cite{yu_evidence_2021, graham_depth-dependent_2024}. Furthermore, in a more general scenario where the mean field $B = \omega/\gamma_\mu$ is non-zero, i.e., in the presence of a unidirectional distributed static local magnetic field, such as static time-reversal symmetry breaking fields. The formula is extended as \cite{noakes_numerical_1986,solt_generalized_1995,wang_quantum_2017}
\begin{equation}
    P_\mathrm{ZF}=\frac{1}{3}+\frac{2}{3}\mathrm{exp}(-\frac{1}{2}\Delta^2t^2)\left[\mathrm{cos}(\omega{t})-(\frac{\Delta^2t}{\omega})\mathrm{sin}(\omega{t})\right]
\end{equation}
Note that for $\mathrm{\Delta/\omega} \gg 1$, the general formula takes the shape of the Gaussian Kubo-Toyabe function. A good distinction between $\mathrm{\Delta}$ and $\mathrm{\omega}$ occurs for $\mathrm{\Delta/\omega} < 0.1$, which is not in this reported case. Given the finite measurement time window up to 9 $\mu$s and $\mathrm{\Delta}$ on the order of 0.2 MHz, the GKT function is therefore an approximation where minusculed changes in $\mathrm{\omega}$ will be absorbed in $\mathrm{\Delta}$ from our fitting procedure. \tcr{Furthermore, the absence of a spontaneous magnetic field can also be demonstrated by fixing the $\Delta$ to be the value at 70 K and allowing $\omega$ to vary as a free fitting parameter. It is observed that for $p{\perp}c$, any $\omega>0$ would increase the residual error of the fitting.}

The muon spin depolarization rate exhibits distinct differences in two orthogonally projected muon spin directions ($p {\parallel} c$ and $p {\parallel} ab$), where $p$ is the projection of the initial muon spin direction. As shown in Fig. 2(c), this behavior closely resembles that of the parent compound \cite{yu_evidence_2021}. Both compounds show an anomaly at $T' = 70$ K, below which the depolarization rates display opposite temperature dependence for different crystallographic directions, i.e., they decrease (increase) along the $ab$-plane ($c$-axis). Given the previously established absence of both magnetism and long-range CDW, the most plausible explanation for the observed muon depolarization is a local lattice distortion, which alters the local nuclear fields at the muon site. A second anomaly occurs below $T" = 35$ K, marking an abrupt enhancement of the muon depolarization rate towards lower temperatures. Its origin is unclear to us, but it has been suggested as a distinct interaction-stabilized long-range electron coherence state \cite{guo_switchable_2022,Guo_long-range_2025}.

\textit{Paramagnetic muon Knight shift- }When the applied external field direction is transverse to the muon spin polarization, the transverse field (TF) \msr\ experiment configuration is established. The corresponding muon spectra was analyzed using a phenomenological depolarization function
\begin{equation}
    P_\mathrm{TF}=\mathrm{exp}(-\frac{1}{2}\mathrm{\Delta_\textrm{TF}}^2t^2)\mathrm{cos}(\omega{t}+\mathrm{\phi}),
\end{equation}
where the oscillatory component of the muon spin polarization in an applied magnetic field is described by a cosine function with an initial phase $\mathrm{\phi}$ of the muon spin. The central frequency $\omega$ is related with the local field $B_\textrm{loc}$ at the muon site by $\omega = \gamma_\mu B_\textrm{loc}$.

In TF-\msr~experiments, the measured difference between the applied field $B_\textrm{ext}$ and $B_\textrm{loc}$ is defined as the muon Knight shift $K$, where $K = (B_\textrm{loc}-B_\textrm{ext})/B_\textrm{ext}$. The non-magnetic semimetals in an applied external magnetic field, $B_\textrm{loc}$ is mainly contributed by the hyperfine interactions related to the spin polarized conduction electrons around the positively charged muons \cite{blundell_longitudinal_2001}. $K$ scales with the local magnetic susceptibility $\chi$, which is proportional to the density of state at the Fermi level \cite{blundell_longitudinal_2001}. For the presented data, $K$ is calculated from $K = [B_\textrm{loc}-B_\textrm{loc}(300\textrm{ K})]/B_\textrm{loc}(300\textrm{ K})$, as the magnetic susceptibility at 300 K is mainly contributed by temperature independent paramagnetisms.

Figure 3(a) reveals an overall negative shift of $K$ upon cooling. This signifies the contact spin polarization at the muon site is antiparallel to the applied field, a specific phenomenon attributed to muon-induced spin polarization in paramagnets \cite{ho_muon_2014}. As the temperature decreases, the trend begins to deviate from the linear behavior, aligning with macroscopic susceptibility measurements shown previously. This deviation is induced by populated spin excited states resulting from the crystal field effect, as illustrated in the inset of Fig. 3(a). The data are qualitatively captured by a two-level model \cite{besara_single_2014} (details in Supplemental Information) with an energy gap $\lambda\approx20$ meV, consistent with local multiplet splitting. \tcr{This value aligns well with \svo, where $\lambda\approx30$ meV for a transition at 100 K \cite{jackeli_magnetically_2009}.}

A Clogston-Jaccarino plot \cite{clogston_interpretation_1964} of $K$ versus $\chi$ is anticipated to exhibit a linear relationship for a paramagnetic contribution \cite{gygax_muon_1980}. As shown in Fig. 3(b), the connection between the high- and low-temperature points follows a linear trend. However, at very low temperatures, a saturation effect can be observed, likely due to the modification of the coupling strength between muons and itinerant electrons in the V $3d-$orbitals \cite{ho_muon_2014}. Meanwhile, the local spin susceptibility measured with \msr~is more sensitive to the anomaly appearing between 20 and 100 K, suggesting the existence of a local effect that is more prominently manifested through local probing techniques, such as \msr.

\begin{figure}[h]
    \centering
    \includegraphics[width=1\linewidth]{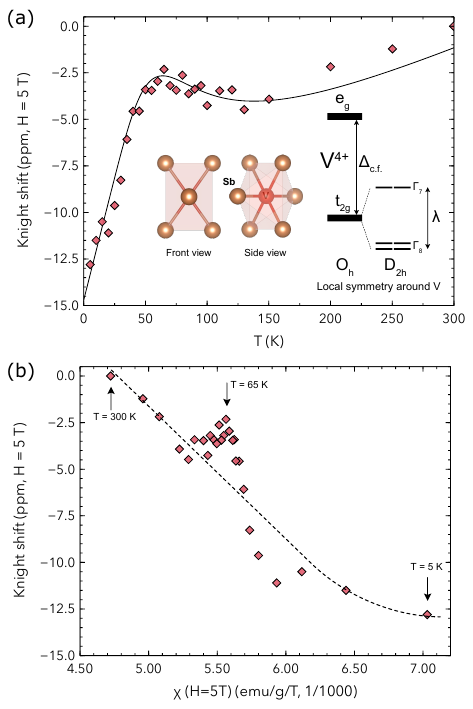}
    \caption{\label{fig:fig4} Results of the muon knight shift experiment: (a) temperature dependence of the muon Knight shifts ($K$) for magnetic fields $H=$ 5 T applied parallel to the $c-$axis ($H{\parallel}c$) of the Ti ($x=0.05$) doped \cvts~single crystal. The solid curve is a simulation of the local magnetic susceptibility of a two-level transition with a gap size of 20 meV. Inset: illustration of the V-Sb octahedron and the V split orbitals in a crystal field. (b) Plot of $K$ vs. the bulk magnetic susceptibility ($\chi$) for $H{\parallel}c$. The dashed curve is a guide to the eyes.
    } 
\end{figure}

\begin{figure*}[ht]
    \centering
    \includegraphics[width=1\linewidth]{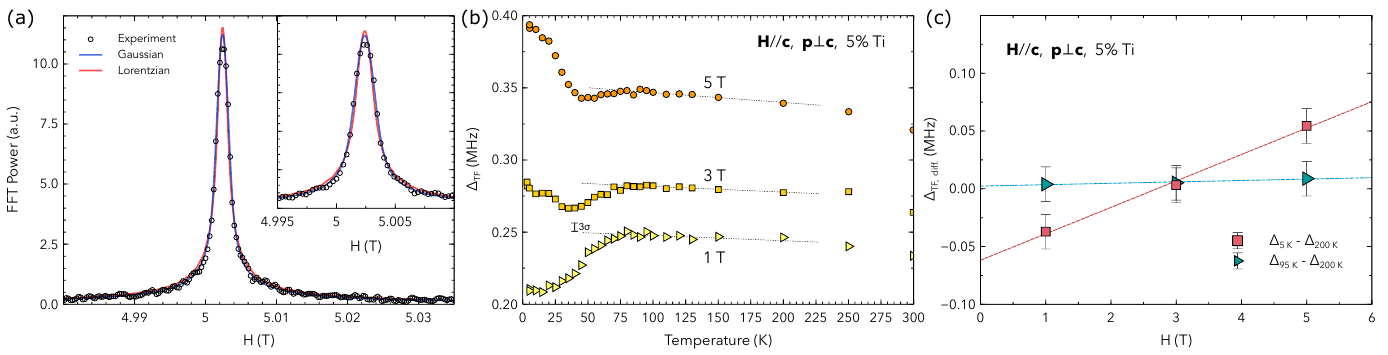}
    \caption{\label{fig:fig5} Magnetic field dependent local magnetic broadening: (a) FFT of the TF muon spectra with $H = 5$ T, the linewidth is fitted using a Gaussian lineshape as depicted in Equ. 3. (b) Temperature-dependent TF-\msr~depolarization rates measured under high magnetic fields between 1 T and 5 T. The horizontal lines are guided to the eyes. (c) Change of the local magnetic field broadening  comparison of the relative changes based on the TF muon depolarization data shown.
    }
\end{figure*}

\textit{Magnetic field induced local field broadening- }
The transverse field muon spin relaxation rate, $\Delta_\textrm{TF}$, can also be obtained from Equ. 3. In general, this rate is a sum of contributions from longitudinal ($1/T_1$) and transverse ($1/T_2$) spin relaxations. Given the absence of notable dynamic relaxation processes, $1/T_1$ is anticipated to be considerably smaller than the $1/T_2$, a supposition supported by the better-defined Gaussian lineshape evident in Fig. 4(a). Consequently, under the applied external transverse magnetic field, the broadening of the local magnetic field remains dominated by $1/T_2$, i.e. static magnetic dipolar interaction.

Remarkably, Fig. 4(b) reveals two distinct regimes in the temperature dependence of the $\Delta_\textrm{TF}$. Above 70 K, $\Delta_\textrm{TF}$ exhibits only a weak temperature dependence. Conversely, below 70 K, a much more pronounced change is observed. This significant variation below 70 K is particularly noteworthy, considering the absence of static magnetic moments arising from electronic magnetic moments, as established by the paramagnetic state discussed earlier. Similar behavior was also reported in the parent \kvs~compound \cite{mielke_time-reversal_2022}, although the underlying cause was attributed to a time-reversal symmetry-breaking charge order.

The $\Delta_\textrm{TF}$ broadening demonstrates a clear field tunability. Fig. 4(c) illustrates that within the low-temperature regime, $\Delta_\textrm{TF}$ scales approximately linearly with the applied external magnetic field at a rate of 0.025 MHz/T. This behavior is unlikely to stem from magnetic field-induced magnetism itself, as there is no corresponding evidence of magnetic susceptibility changes up to 7 T. Consequently, other mechanisms must be responsible for this field-dependent broadening.

One potential origin for this phenomenon is inhomogeneous broadening associated with the second moment (variance) of the nuclear dipolar field distribution \cite{schenck_modelling_1997}. This effect can be significantly enhanced by factors that break local symmetry, such as non-cubic local lattice symmetry or an inhomogeneous demagnetization field arising from the anisotropy in the local spin density of the surrounding electron cloud. While the contribution of the Fermi contact field usually remains isotropic \cite{onuorah_muon_2018}, muon spin conserves the initial polarization in electrical dipolar interaction, the observed result is due to a modification of the distance between the muon and their nearby nucleus. This modification causes slightly altered local magnetic dipolar field strength or a change in the Fermi contact strength. Crucially, both scenarios imply a lower crystalline environment for the muon site \cite{solt_generalized_1995, huang_calculation_2012, le_yaouanc_muon_2010}. The enhanced significance of this effect at lower temperatures suggests a connection to orbital dynamics. Specifically, orbital excitations prevalent at higher temperatures tend to average out distortions, leading to a higher effective symmetry of the crystal lattice. In contrast, the low-temperature ground state exhibits greater distortion, likely stemming from particular orbital occupations that establish an anisotropic electronic configuration. This ground state distortion may arise from orbital ordering within the lattice \cite{Huang_Revealing_2025}, driven by finite spin-orbit coupling. This demonstrates intricate lattice-orbital entanglement governing local magnetic response, even without strong electronic correlations.

Although orbital order in kagome superconductors, such as \cvs~\cite{song_orbital_2022}, has been primarily associated with charge ordered states, exemplified by star-of-David distortions in undoped compounds. Its emergence is therefore regarded linked to the CDW transition temperature. Importantly, this study establishes that orbital excitations, and potentially emergent orbital order, persist in the doped compound despite the complete suppression of long-range CDW order. \tcr{The emergence of a hidden phase, possibly related to octupolar ordering in $d$-electron systems, could be attributed to the orbital ordering of low-energy Kramers doublets in $d^1$ ions} \cite{jackeli_magnetically_2009}. Recently, a more peculiar electronic state has also been proposed, involving a staggered spin-flux coupled with the breathing mode of lattice distortion, potentially present across all phonon frequencies smaller than the $\lambda$~\cite{zhang_spontaneous_2025}. This demonstrates a previously underrated, universal decoupling of orbital physics from charge ordering, with consistent ZF- and TF-\msr~measurements confirming that local lattice distortions drive the observed phenomena.

\textit{Conclusion- }
Our investigations corroborate that the Ti-doped \cvts~compound exhibits paramagnetism at both local and macroscopic levels, with no spontaneous magnetic fields detected. It is noteworthy that orbital excitations predominantly influence the local magnetic susceptibility, even in the absence of charge order. This dominance constitutes a significant intrinsic effect inherent to the layered V-Sb crystal structure. Furthermore, an atypical ZF muon spin relaxation response is observed, manifesting as an anisotropic nuclear local magnetic field distribution at the muon site. This anomaly persists despite the absence of zero- and field-induced magnetism in macroscopic susceptibility measurements up to 7 T, suggesting its unconventional origin and independence from electron spin magnetism. Notably, the field broadening in TF-\msr~experiments aligns with an anisotropic local field distribution, indicating a scenario driven by temperature-dependent lattice distortions, potentially associated with orbital ordering at low temperatures. Significantly, similar ZF-\msr~behavior is observed in the parent compound, establishing orbital excitations as an inherent property of the layered V-Sb structure with potential universality across kagome \avs~compounds. These excitations exhibit strong coupling to the lattice via spin-orbital coupling, providing a means to enable exotic lattice dynamics, including circularly polarized phonons, in nominally non-chiral crystal lattices \cite{juraschek_giant_2022,chaudhary_giant_2024}.

\vspace{10pt}

\vspace{10pt}
\textit{Acknowledgements- }
We would like to thank the discussion with Michael Graf, Stephen Wilson, Claude Monney, Zhongxian Zhao, and the help of Li Yu at the early stage of this study. We acknowledge the allocation of muon beamtimes and the technical support from Hubertus Luetkens, Zurab Guguchia, Robert Scheuermann at the GPS and the HAL beamline of S{$\mu$}S. The work at IOP is supported by National Key Research and Development Program of China (Grant Nos. 2022YFA1204100, 2022YFA1403903, and 2023YFA1406101) and the National Natural Science Foundation of China (62488201). Z.W. is supported by the U.S. Department of Energy, Basic Energy Sciences Grant DE-FG02-99ER45747.

\bibliographystyle{naturemag}
\bibliography{exported_v2}

\end{document}